\def\Camera{0}%
\def\Camera{1}%
\def\Draft{1}% draft=1 or no draft = 0
\def\Draft{0}% draft=1 or no draft = 0
\def\Supp{0}% does not include supp material
\def\Supp{1}% includes supp material
\newcommand{\Author}{Perrinet}%
\newcommand{\FirstName}{Laurent U.}%
\newcommand{\Institute}{Institut de Neurosciences de la Timone (UMR7289)\\ CNRS / Aix-Marseille Universit\'e}%
\newcommand{\Address}{27, Bd. Jean Moulin, 13385 Marseille Cedex 5, France}%
\newcommand{\Website}{http://invibe.net/LaurentPerrinet/Publications/Perrinet16EUVIP}%
\newcommand{\Email}{Laurent.Perrinet@univ-amu.fr}%
\newcommand{\Title}{Biologically-inspired characterization of sparseness in natural images}%
\newcommand{\Abstract}{
Natural images follow statistics inherited by the structure of our physical (visual) environment. In particular, a prominent facet of this structure is that images can be described by a relatively sparse number of features. We designed a sparse coding algorithm biologically-inspired by the architecture of the primary visual cortex. We show here that coefficients of this representation exhibit a heavy-tailed distribution. For each image, the parameters of this distribution characterize sparseness and vary from image to image. To investigate the role of this sparseness, we designed a new class of random textured stimuli with a controlled sparseness value inspired by our measurements on natural images. Then, we provide with a method to synthesize random textures images with a given statistics for sparseness that matches that of some given class of natural images and provide perspectives for their use in neurophysiology.
}
\newcommand{\Keywords}{Machine vision, Image texture, Computer vision, Neuroscience}%
\newcommand{\Acknowledgments}{%
L.U.P.\ was supported by ANR project "BalaV1" ANR-13-BSV4-0014-02. Correspondence and requests for materials should be addressed to LUP (email:\Email ). Code and supplementary material available at \url{\Website/Publications/Perrinet16EUVIP}.
} %

\documentclass[a4paper, 10pt, twocolumns]{article}
\usepackage{amsmath,epsfig}%spconf,
\usepackage[numbers,comma,sort&compress]{natbib} % ,round
\usepackage{tikz}
\DeclareGraphicsExtensions{.pdf,.png,.jpg}
\graphicspath{{../test/results/},{./}}% 

\usepackage{amsmath}
\usepackage{amssymb}
\usepackage{amsthm}

\newcommand{\dotp}[2]{\langle #1,\,#2\rangle}

\newcommand{\eql}[1]{\begin{equation}#1\end{equation}}
\usepackage{siunitx}

\newcommand{\RR}{\ensuremath{\mathbb{R}}}

\newcommand{\NN}{\ensuremath{\mathbb{N}}}

\if 1\Draft
\usepackage{setspace}
\fi

\title{\Title}
\author{%
\FirstName\ \Author\thanks{See {\Website}.}\\
\Institute\ \\ \Address\ \\
\texttt{\Email} \\
}
\if\Camera1
\newcommand{\url}[1]{{\rm #1}}
\else
\usepackage[unicode,linkcolor=blue,citecolor=blue,filecolor=black,urlcolor=blue,pdfborder={0 0 0}]{hyperref}%
\hypersetup{%
pdftitle={\Title},%
pdfauthor={\Author < \Email > \Address - \Website },%
pdfkeywords={\Keywords},%
pdfsubject={\Acknowledgments}%
}%
\fi
\begin{document}
\if\Draft1
\doublespacing
\fi

%\ninept
%
\maketitle
\begin{abstract}
\Abstract
\end{abstract}
%
%\begin{keywords}
%\Keywords
%\end{keywords}
%%
\paragraph{BibTex entry}~~\\
% to update from http://invibe.net/cgi-bin/index.cgi/Publications/Ravello17droplets?action=raw
\begin{verbatim}
@inproceedings{Perrinet16EUVIP,
author = {Perrinet, Laurent U.},
booktitle = {2016 6th European Workshop on Visual Information Processing (EUVIP)},
doi = {10.1109/EUVIP.2016.7764592},
isbn = {978-1-5090-2781-1},
month = {oct},
pages = {1--6},
publisher = {IEEE},
title = {Biologically-inspired characterization of sparseness in natural images},
url = {http://ieeexplore.ieee.org/document/7764592/},
year = {2016}
}
\end{verbatim}
%⎯⎯⎯⎯⎯⎯⎯⎯⎯⎯⎯⎯⎯⎯⎯⎯⎯⎯⎯⎯⎯⎯⎯⎯⎯⎯⎯⎯⎯⎯⎯⎯⎯⎯⎯⎯⎯⎯⎯⎯⎯⎯⎯⎯⎯⎯⎯⎯⎯⎯⎯⎯⎯⎯⎯⎯⎯⎯⎯⎯⎯⎯⎯⎯⎯⎯⎯⎯⎯⎯⎯⎯⎯⎯⎯⎯⎯⎯⎯⎯⎯⎯
\section{Motivation}
%⎯⎯⎯⎯⎯⎯⎯⎯⎯⎯⎯⎯⎯⎯⎯⎯⎯⎯⎯⎯⎯⎯⎯⎯⎯⎯⎯⎯⎯⎯⎯⎯⎯⎯⎯⎯⎯⎯⎯⎯⎯⎯⎯⎯⎯⎯⎯⎯⎯⎯⎯⎯⎯⎯⎯⎯⎯⎯⎯⎯⎯⎯⎯⎯⎯⎯⎯⎯⎯⎯⎯⎯⎯⎯⎯⎯⎯⎯⎯⎯⎯⎯
%------------------------------%
%: see Figure~\ref{fig:EUVIP_lena}
\begin{figure}[ht!]%[p!]
\centering{\includegraphics[width=.7\linewidth]{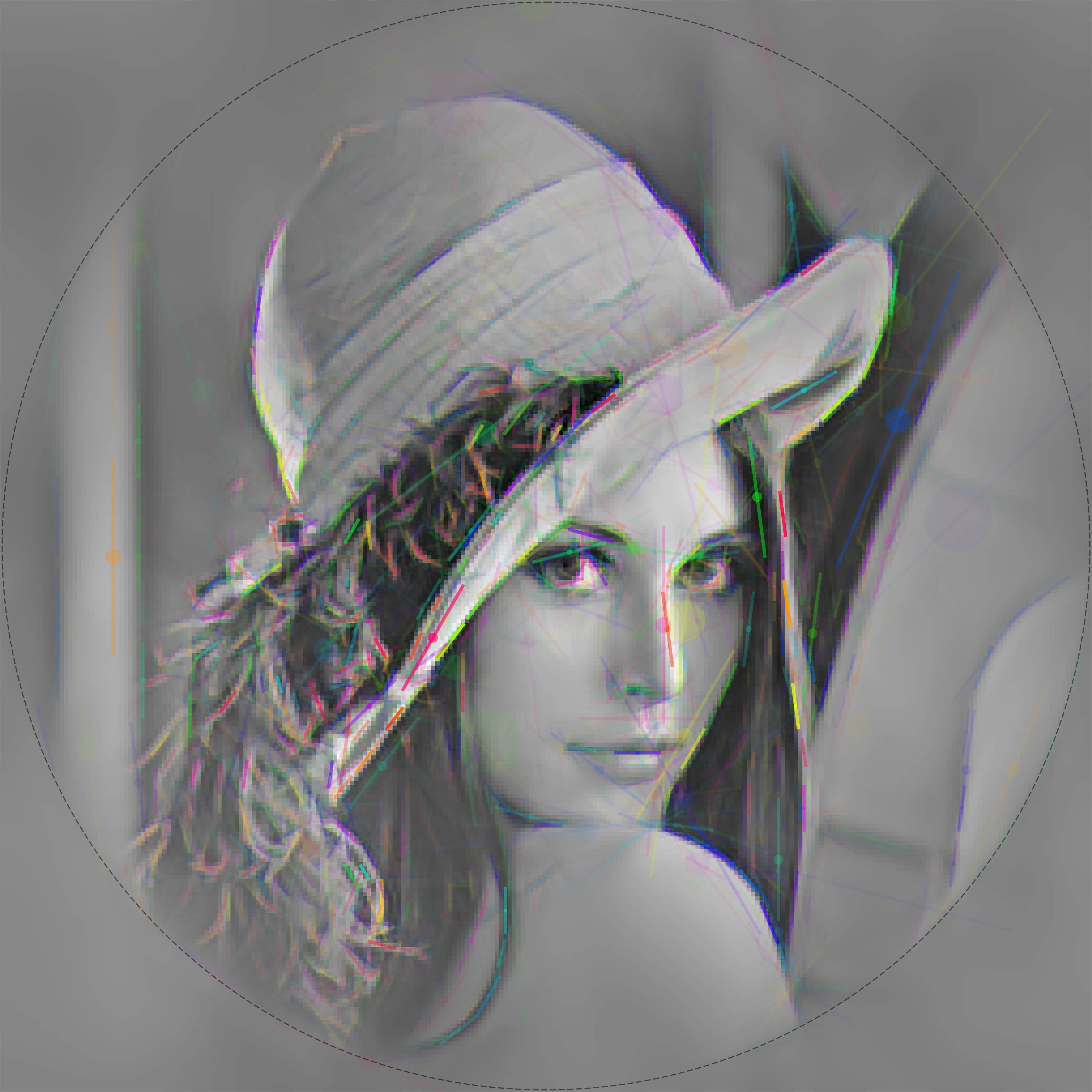}}
\caption{
{\bf Sparse coding of natural images using log-Gabor edges and applied to ``Lena''}:
An instance of the reconstruction of a natural image as a sum of elementary edges using sparse coding. Parameters for each edge are its position, orientation, scale and scalar amplitude of the coefficient. Edges are overlaid on this reconstructed image as colored segments, length denoting scale and hue the orientation\if\Supp1\ --- see SI Section~\ref{sec:sparse} for a full description of the algorithm\fi. Edges outside the dashed circle are discarded to avoid artifacts. Few edges (here $1024$) are necessary to efficiently represent the image: the distribution of features is \emph{sparse}. The position and orientation of the set of edges gives information about the shape of visual objects, while the distribution of coefficients characterizes the texture content of the image.
\label{fig:EUVIP_lena}}%
\end{figure}
\begin{figure*}[t!]%[p!]
\centering{
\begin{tikzpicture}
\draw [anchor=north west] (.0\linewidth, 0) node {\includegraphics[width=.5\linewidth]{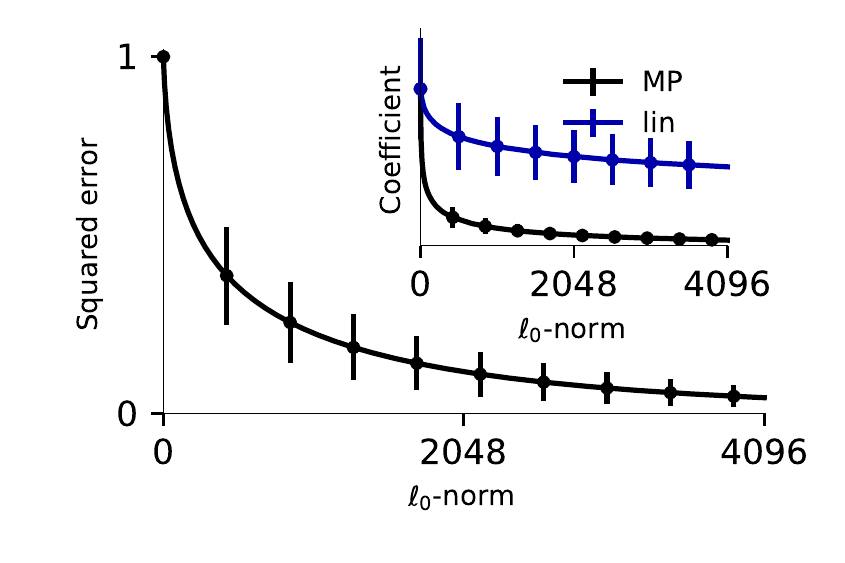}};
\draw [anchor=north west] (.5\linewidth, 0) node {\includegraphics[width=.5\linewidth]{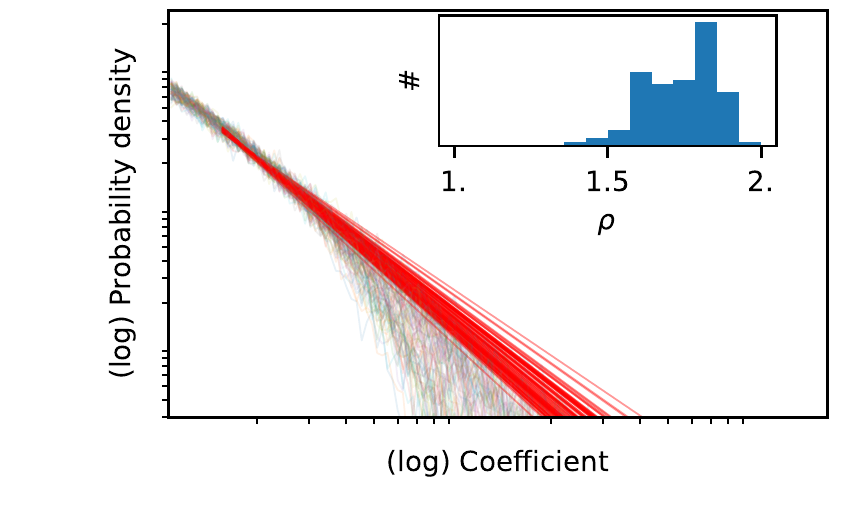}};
\draw (.00\linewidth, 0) node [above right=0mm] {$\mathsf{A}$};
\draw (.5\linewidth, 0) node [above right=0mm] {$\mathsf{B}$};
\end{tikzpicture}}
\caption{
{\bf Coefficients from the Sparse coding of images in the primary visual cortex (V1) follows regular statistics}: \textsf{(A-inset)}~Coefficients decrease as a function of their rank (average across images $\pm$ standard deviation). This decrease is faster when using a sparse coding mechanism (Matching Pursuit, 'MP') than when just ordering the linear coefficients ('lin'). \textsf{(A)}~For the sparse coding, we controlled the quality of the reconstruction from the edge information such that the residual energy is less than $5\%$ over the whole set of images, a criterion met on average when identifying $2048$ edges per image for images of size $256\times 256$ (that is, a relative sparseness of $\approx 0.02\%$ of activated coefficients).
\textsf{(B)}~When observing on an image-by-image basis the histogram of the coefficients' amplitude, each follows a generic log-Normal probability density function, only controlled by two parameters (the mode and bandwidth $B$) for each of natural image. While the mode parameter did not vary significantly, higher values of $B$ qualitatively correspond to sparser structures.%\if\Supp1 (Supplementary Figure~\ref{fig:EUVIP_proba_scaled} shows that once scaled, these distribution align almost perfectly)\fi.
\textsf{(B-inset)}~The distribution of these sparseness parameters shows a variety of values from the most sparsely distributed (right, $B\approx1.$) to those containing mostly dense textures (left, $B\approx3.$).
\label{fig:EUVIP_sparseness}}%
\end{figure*}

%
%------------------------------%
% sparseness in sparse coding
Natural images, that is, visual scenes that are relevant for an animal, most generally consist of the composition of visual objects, from vast textured backgrounds to single isolated items. By the nature of the structure of the physical (visual) world, these visual items are often sparsely distributed, such that these elements are clustered and a large portion of the space is void: A typical image thus contains vast areas which are containing little information while most information is concentrated in a small portion of the image (see Figure~\ref{fig:EUVIP_lena}\if\Supp1 and Supplementary Figure~\ref{fig:EUVIP_lena_movie} \fi). A first formalism to quantify this sparseness is to consider the distribution of coefficients obtained by coding such images using a bank of filters resembling the edge-like profiles observed in the primary visual cortex (V1) of primates\if\Supp1~(See Supplementary Figure~\ref{fig:EUVIP_loggabor})\fi. From this linear representation, it is possible to determine a near-to-optimal coding formalism which involves the inversion of a model of the (forward) neural transformation~\citep{Simoncelli2004}.  As a matter of fact, a normative explanation for the coding of images into activity in the neural tissue is the optimal representation of visual information~\citep{Atick92}. Inspired by the representation in V1, a popular method requires the efficient modeling of images as the sparsest combination of edges and such a coding strategy indeed resembles the neural activity (that is, the spike patterns) observed in V1~\citep{Perrinet15bicv}. Using such a method, it was previously shown that any image from a database of static, grayscale natural images may be coded using solely a relatively few number of coefficients (See Figure~\ref{fig:EUVIP_sparseness}\if\Supp1 and SI Section~\ref{sec:sparse} for a full description of the algorithm\fi). Crucially, applying this sparse coding method on this database, we observed that some images were sparser than others but that globally they all fitted well a similar prototypical probability distribution function (pdf). The parameters of this pdf provide with a quantitative descriptor for sparseness and is characteristic for this property. However, it is largely unclear how this property inherent to the structure of natural images may be used in the early visual system or in computer vision. %

% efficiency
The different levels of sparseness observed in images are qualitatively linked to their textural content. Indeed, different levels of roughness in any image correspond to different distributions of the coefficients corresponding to its representation. Such an analysis was previously performed for instance by characterizing the Hurst parameter for two-dimensional fractional Brownian motion~\citep{Kaplan1999}. However, this analysis was done on the linear coefficients of Gabor features, while we rather wish here to characterize sparseness in the above-mentioned biologically inspired generative model for natural images. Compared to a linear representation, this coding is non-linear and the distribution of edges' parameters are \emph{a priori} different from that described in~\citep{Kaplan1999}. In particular, while the distribution of position, scale and orientation are well characterized, an unified theory for the distribution of sparse coefficients is still lacking. As such, a major challenge is to refine the characterization of sparseness in natural images such that it conforms to the widest variety of visual scenes. Here, by using a bio-inspired representation based on the architecture of the primary visual cortex, we will first characterize the pdf of sparse coefficients on an image-by-image basis and then propose a simple characterization of sparseness based on the number of non-zero coefficients that we apply to texture synthesis.

% Contributions / outline
Our first contribution is to show that while the distribution of coefficients in natural images is stereotyped, the underlying parameters display some variability (see Figure~\ref{fig:EUVIP_sparseness}). Our second contribution is an efficient extension of a texture synthesis model which allows to parameterize the sparseness in the texture. This formulation is important to challenge the effects of the model's parameters in future neurophysiological experiments. Indeed, in neurophysiology, a popular technique to challenge the coding hypothesis is to test the system (V1) using complex stimuli in the form of stochastic stimuli~\citep{Touryan01}. These stimuli correspond to instances of the generative model for natural images which most likely describe a wide range of image instances. From that perspective, we motivate the generation of an optimal stimulation within a stationary Gaussian dynamic texture model (see Figure~\ref{fig:DropLets}). We base our model on a previously defined heuristic coined ``Motion Clouds''~\citep{Leon12}. These contributions show that overall, sparseness in natural images is an important parameter to understand natural scenes. Beyond the contribution to the understanding of the models underlying image representation which may have applications to tune computer vision algorithms, this may have applications outside computer vision, for instance for the optimal stimulation of neurophysiological settings as for instance in a retinal implant.
%------------------------------%
%: see Figure~\ref{fig:EUVIP_sparseness}
%------------------------------%
%⎯⎯⎯⎯⎯⎯⎯⎯⎯⎯⎯⎯⎯⎯⎯⎯⎯⎯⎯⎯⎯⎯⎯⎯⎯⎯⎯⎯⎯⎯⎯⎯⎯⎯⎯⎯⎯⎯⎯⎯⎯⎯⎯⎯⎯⎯⎯⎯⎯⎯⎯⎯⎯⎯⎯⎯⎯⎯⎯⎯⎯⎯⎯⎯⎯⎯⎯⎯⎯⎯⎯⎯⎯⎯⎯⎯⎯⎯⎯⎯⎯⎯
\section{Biologically-inspired sparse coding}
%⎯⎯⎯⎯⎯⎯⎯⎯⎯⎯⎯⎯⎯⎯⎯⎯⎯⎯⎯⎯⎯⎯⎯⎯⎯⎯⎯⎯⎯⎯⎯⎯⎯⎯⎯⎯⎯⎯⎯⎯⎯⎯⎯⎯⎯⎯⎯⎯⎯⎯⎯⎯⎯⎯⎯⎯⎯⎯⎯⎯⎯⎯⎯⎯⎯⎯⎯⎯⎯⎯⎯⎯⎯⎯⎯⎯⎯⎯⎯⎯⎯⎯
Let us first describe the method used for efficiently extracting edges in natural images before measuring the statistics of the edges' parameters. The first step of our method involves defining the dictionary of templates (or filters) for detecting edges. In particular, we use a synthesis model for the edge representation, so that the edges we detect are guaranteed to be sufficient to regenerate the image with a low error. For that, we use a log-Gabor representation, which is well suited to represent a wide range of natural images~\citep{Fischer07}. This representation gives a generic model of edges parameterized by their shape, orientation, and scale. We set the range of these parameters to match what has been reported for the responses in primates' V1. In particular, we set the bandwidth of the Fourier representation of the filters to $1$ and $\pi/8$ respectively in log-frequency and polar coordinates to get a family of elongated and thus orientation-selective filters (see~\citep{Fischer07cv}\if\Supp1 , see Supplementary Figure~\ref{fig:EUVIP_loggabor}\fi). Prior to the analysis of each image, we used the spectral whitening filter~\citep{Olshausen97} to provide a good balance of the energy of output coefficients~\citep{Perrinet03ieee,Fischer07}.

Although orthogonal representations are popular for computer vision due to their computational tractability, it is desirable in our context that we have a high over-completeness in the representation. Indeed, the representation of edges, similar to the receptive fields in the primary visual cortex, should provide with a relative invariance to common geometrical transformations (translation, rotation, and scale). First, a linear convolution model automatically provides a translation-invariant representation. Such invariance can be extended to rotations and scalings by choosing to multiplex these sets of filters at different orientations and spatial scales. Ideally, the parameters of edges would vary in a continuous fashion, to provide relative invariance. For the set of $256\times 256$ images, we chose to have $8$ dyadic levels (that is, doubling the scale at each level) with $24$ different orientations. Orientations are measured as an undirected angle in radians, in the range from $0$ to $\pi$ (but not including $\pi$). Tests with a range of different numbers of orientations and scales yielded similar results~\citep{Perrinet15bicv}. Finally, each image is transformed into a pyramid of coefficients. This pyramid consists of approximately $4/3\times256^{2}\approx8.7\times10^4$ pixels multiplexed on $8$ scales and $24$ orientations, that is, approximately $16.7\times10^6$ coefficients, an over-completeness factor of about $256$.

This transform is linear and can be performed by a simple convolution repeated for every edge type. Following~\citep{Fischer07cv}, convolutions were performed in the Fourier (frequency) domain for computational efficiency. The Fourier transform allows for a convenient definition of the edge filter characteristics, and convolution in the spatial domain is equivalent to a simple multiplication in the frequency domain. By multiplying the envelope of the filter and the Fourier transform of the image, one may obtain a filtered spectral image that may be converted to a filtered spatial image using the inverse Fourier transform. We exploited the fact that by omitting the symmetrical lobe of the envelope of the filter in the frequency domain, the output of this procedure gives a complex number whose real part corresponds to the response to the symmetrical part of the edge, while the imaginary part corresponds to the asymmetrical part of the edge (see~\citep{Fischer07cv} for more details). More generally, the modulus of this complex number gives the energy response to the edge (comparable to the response of complex cells in area V1), while its argument gives the exact phase. This property further expands the richness of the representation. Overall, such a representation is implemented in the \verb+LogGabor+ package of Python scripts\footnote{Available at \url{https://github.com/bicv/LogGabor} and documented at \url{https://pythonhosted.org/LogGabor}.}.

Because this dictionary of edge filters is over-complete, the linear representation would give an inefficient representation of the distribution of edges (and thus of edge co-occurrences) due to {\it a  priori} correlations between coefficients. Therefore, starting from this linear representation, we searched for the most sparse representation. Minimizing the $\ell_0$ pseudo-norm (the number of non-zero coefficients) leads to an expensive combinatorial search with regard to the dimension of the dictionary (it is NP-hard). As proposed first by~\citep{Perrinet02sparse}, we may approximate a solution to this problem using a greedy approach, namely the Matching Pursuit algorithm. This class of algorithms gives a generic and efficient representation of edges, as illustrated by the example in Figure~\ref{fig:EUVIP_lena}. For the simulations presented here, sparse coding of images is implemented in the \verb+SparseEdges+ package of Python scripts\footnote{These scripts are available at \url{https://github.com/bicv/SparseEdges} and documented at \url{https://pythonhosted.org/SparseEdges}.} which depend on packages NumPy (version 1.11.1) and SciPy (version 0.18.0)~\citep{Oliphant07} on a cluster of Linux computing nodes. Visualization was performed using Matplotlib (version 1.5.1)~\citep{Hunter07}. Our goal is to study how the statistics of edges' occurrences vary across a set of natural images. In order to perform that, wes first defined a database of $600$ images\footnote{Publicly available at \url{http://cbcl.mit.edu/software-datasets/serre/SerreOlivaPoggioPNAS07}.}. The performance of the algorithm can be measured quantitatively by reconstructing the image from the list of extracted edges. Measuring the ratio of extracted energy in the images, $N=2048$ edges were enough to extract on average more than $95\%$ of the energy of $256\times 256$ images on all sets of images (see Figure~\ref{fig:EUVIP_sparseness}-A). Then, we tested different hypothesis for the distribution of coefficients and tested the synthesis of textures using this generative model well tuned to these natural images. All these different steps are reproducible as a set of notebooks.%
%⎯⎯⎯⎯⎯⎯⎯⎯⎯⎯⎯⎯⎯⎯⎯⎯⎯⎯⎯⎯⎯⎯⎯⎯⎯⎯⎯⎯⎯⎯⎯⎯⎯⎯⎯⎯⎯⎯⎯⎯⎯⎯⎯⎯⎯⎯⎯⎯⎯⎯⎯⎯⎯⎯⎯⎯⎯⎯⎯⎯⎯⎯⎯⎯⎯⎯⎯⎯⎯⎯⎯⎯⎯⎯⎯⎯⎯⎯⎯⎯⎯⎯
\section{Distribution of sparse coefficients and the design of the ``DropLets'' stimuli}
%⎯⎯⎯⎯⎯⎯⎯⎯⎯⎯⎯⎯⎯⎯⎯⎯⎯⎯⎯⎯⎯⎯⎯⎯⎯⎯⎯⎯⎯⎯⎯⎯⎯⎯⎯⎯⎯⎯⎯⎯⎯⎯⎯⎯⎯⎯⎯⎯⎯⎯⎯⎯⎯⎯⎯⎯⎯⎯⎯⎯⎯⎯⎯⎯⎯⎯⎯⎯⎯⎯⎯⎯⎯⎯⎯⎯⎯⎯⎯⎯⎯⎯
Empirically, we observed that the distribution of sparse coefficients for any natural image fits with a log-Normal probability distribution function (see Figure~\ref{fig:EUVIP_sparseness}-B). In particular, the histogram of occurrences of the absolute sparse coefficients for any given image follows a smooth curve in a log-log plot, suggesting a fit with a log-Nornal distribution as parameterized by the mode and bandwidth $B$. We fitted each histogram using the Levenberg-Marquardt method from SciPy~\citep{Oliphant07}. In particular, this was validated by checking the goodness of fit for each image. While the mode did not vary significantly across the database of natural images, we observed that the bandwidth $B$ of the distribution was significantly different across the database. This is shown as the histogram of the $B$ values obtained for each image (see Figure~\ref{fig:EUVIP_sparseness}-B, inset). This demonstrates that natural images show a wide range of sparseness in their coefficients from very sparse ($B\approx 3.$) to more dense ($B\approx 1.$). Qualitatively, this corresponds to images which are containing respectively less to more dense textures. An interpretation for that class of distributions comes from the fact that such pdf emerges from the mixing of different sources with narrow bandwidths: a denser mixing produces a broader bandwidth. As a consequence, sparseness could be simply parameterized as the intensity of the mixing of such events, or put even more simply, to the number of non-zero coefficients. Such knowledge can now be injected in a model for texture synthesis.

%------------------------------%
%: see Figure~\ref{fig:DropLets}
\begin{figure*}[ht!]%[p!]
\centering{
\includegraphics[width=.99\linewidth]{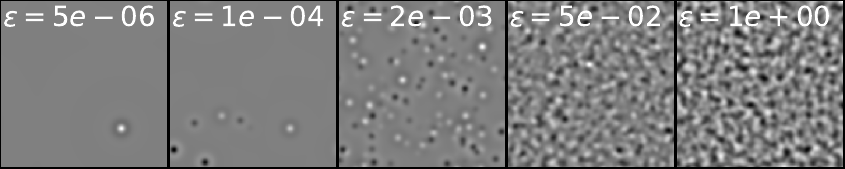}
}
\caption{
{\bf DropLets stimuli:} We design DropLets as random phase textures where, similarly to the variety observed in a set of natural images, the distribution of coefficients is parameterized by different levels of sparseness. From the observations in a range of natural images (see Figure~\ref{fig:EUVIP_sparseness}-B, inset), we define the sparseness $\epsilon$ as the relative number of non-zero coefficients, from very sparse ($\epsilon=5e-06$) to dense ($\epsilon=1$), and with $\epsilon$ values chosen on a geometrical scale between those extremes. Note that, while for $\epsilon=1$ these textures are fully dense and correspond to a linearly filtered noise,  other values correspond to progressively sparser textures as $\epsilon$ tends to lower values. Moreover, while the textons used to synthesize these textures are arbitrary in our formalism, targetting a neurophysiological experiment on the retina, we have chosen here for the sake of simplicity to use symmetrical Mexican-hat profiles (``drops''\if\Supp1 , see Supplementary Figure~\ref{fig:EUVIP_loggabor}\fi) at a single scale which is optimal for retinal ganglion cells.
\label{fig:DropLets}}%
\end{figure*}%
%------------------------------%
From these observations, we propose a mathematically-sound derivation of a general parametric model of synthetic textures. This model is defined by aggregation, through summation, of a basic spatial ``texton'' template $\phi(x, y)$. The summation reflects a transparency hypothesis, which has been adopted for instance in~\citep{Galerne11}. In particular, this simple generative model considers independent, transparent elementary features. While one could argue that this hypothesis is overly simplistic and does not model occlusions, it leads to a tractable framework of stationary Gaussian textures, which has proved useful to model static micro-textures~\citep{Galerne11} and dynamic natural phenomena~\citep{2014-xia-siims}. In particular, the simplicity of this framework allows for a fine tuning of frequency-based (Fourier) parameterization, which is desirable for the interpretation of neuro-physiological experiments. In summary, it states that luminance $I(x)$ for $x \in \RR^2$ is defined as a random field~\citep{Vacher15}: \eql{\label{eq-deadleaves}
I (x) = \sum_{i \in \NN} a_i \cdot \phi(x-x_i)} where the respective random variables parameterize respectively each texton's scalar value $a_i$ and position $x_i \in \RR^2$. As was previously mentioned~\citep{Leon12,Vacher15}, this defines the set of ``Motion Clouds'' stimuli as that defined by: \eql{\label{eq-MC} I (x) = [\sum_{i \in \NN} a_i \cdot\delta(x-x_i)] \ast \phi(x) } where $\ast$ denotes the convolution operator over variable $x$. Noting the Fourier transform as $\mathcal{F}$ and $f \in \RR^2$ the frequency variables, the image $I$ is thus a stationary Gaussian random field of covariance having the power-spectrum $\mathcal{E}=\mathcal{F}(\hat{\phi})$ (for a proof, see~\citep{Vacher15}). It comes
\begin{equation*}
  \mathcal{F}(I) (f) = A \cdot e^{i \cdot \Phi} \cdot \mathcal{E} (f)
\end{equation*}
where
\begin{equation*}
A \cdot e^{i \cdot \Phi} = \sum_{i\in\NN} a_i \cdot e^{-2i\pi \dotp{f}{x_i}}
\end{equation*}
corresponds to an iid random phase field scaled by $A \in \RR^+$. Such a random texture is parameterized by the (positive-, real-valued) envelope $\mathcal{E}$. To match the statistics of natural scenes or some category of textures, the envelope $\mathcal{E}$ is usually defined as some average spatio-temporal coupling over a set of natural images. Note that it is possible to consider any arbitrary texton $\mathcal{F}^{-1}(\mathcal{E})$, which would give rise to more complicated parameterizations for the power spectrum. In particular, we may use the same set of edge-like filters used in the sparse coding algorithm above.

The originality provided here is to introduce an explicit sparsity in equation~\ref{eq-MC} by parameterizing a sparse distribution for the coefficients $a_i$. Mathematically, the $a_i$ are drawn from a heavy-tailed probability distribution function $p_a$ such as that found in natural images (See Figure~\ref{fig:EUVIP_sparseness}-B). Such assumptions were previously used for generating procedural noise in computer vision~\citep{Lagae09}, but we focus here on a definition based on a model of sparseness in images. A similar endeavor was initiated by~\citep{Sallee03} by defining a mixture of a Dirac and a Gaussian distributions, but here, we derive it from the sparse coding observed in a set of natural images. First, let's define $E=\sum_{i} a_i \cdot\delta(x-x_i)$ the (sparse) matrix corresponding to the textons' coefficients at different positions. We also note that we can still compute the image in Fourier space, but with a random phase field which by linearity of $\mathcal{F}$ is simply given $\forall f$ by : \eql{\label{eq-phase}   \mathcal{F}(I) (f) = \mathcal{F}(E)(f) \cdot \mathcal{E} (f)} Compared to the pure summation, a first advantage of this procedure is that the computation time does not depend on the number of events. A further advantage of equation~\ref{eq-phase} in generating these stimuli is that for any given instance of the noise, we know the position $(x, y)$ of each event. When using rotationally symmetric, Mexican-hat profiles, the resulting texture resembles the shot noise image of drops of rain on water, such that we coined this set of stimulus as ``DropLets'' (See Figure~\ref{fig:DropLets}). In all generality, for any given texton (for instance a choice of a given orientation $\theta$ and scale $\sigma$), the implementation of this texture generation is of the same order of complexity as that of Motion Clouds and allows for the easy control of the sparseness in the stimulus. These different (uncoherent) ``layers'' may then be summed to form a texture of sparsely distributed edge-like features.
%⎯⎯⎯⎯⎯⎯⎯⎯⎯⎯⎯⎯⎯⎯⎯⎯⎯⎯⎯⎯⎯⎯⎯⎯⎯⎯⎯⎯⎯⎯⎯⎯⎯⎯⎯⎯⎯⎯⎯⎯⎯⎯⎯⎯⎯⎯⎯⎯⎯⎯⎯⎯⎯⎯⎯⎯⎯⎯⎯⎯⎯⎯⎯⎯⎯⎯⎯⎯⎯⎯⎯⎯⎯⎯⎯⎯⎯⎯⎯⎯⎯⎯
\section{Perspectives in neurophysiology}
%⎯⎯⎯⎯⎯⎯⎯⎯⎯⎯⎯⎯⎯⎯⎯⎯⎯⎯⎯⎯⎯⎯⎯⎯⎯⎯⎯⎯⎯⎯⎯⎯⎯⎯⎯⎯⎯⎯⎯⎯⎯⎯⎯⎯⎯⎯⎯⎯⎯⎯⎯⎯⎯⎯⎯⎯⎯⎯⎯⎯⎯⎯⎯⎯⎯⎯⎯⎯⎯⎯⎯⎯⎯⎯⎯⎯⎯⎯⎯⎯⎯⎯
In this paper, we have characterized the different levels of sparseness which are present in natural images and synthesized random textures with parameterized levels of sparseness. First, we have replicated the observation that natural images follow a prototypical structure for the probability density function of the coefficients that characterize them. Importantly, we have shown that on an image-by-image basis, this structure is qualitatively well captured by using the number of non-zero coefficients as a measure of the sparseness of a given image, from a dense texture to a highly sparse configuration. Based on these results, we designed random textures which replicate this parameterization of their sparseness.

In a recent experimental project using such a set of textures, it has been shown that the neural code as recorded in the retina responds differentially to these different levels of sparseness~\citep{Ravello2015}. In particular, we used the set of stimuli (See Figure~\ref{fig:DropLets}) to evaluate the response on ganglion cells on an \emph{ex vivo} preparation and analyzed the response of the population of neurons as a function of the sparseness parameter. Such analysis proved that the retinal neural code showed differential responses with respect to the sparseness of images and preliminary measures on efficiency suggest that it is tuned for a given level of sparseness: In that experimental results, we have shown that there is a limit in sparseness for which the retina can respond optimally; beyond that point, the response is more or less the same, meaning that the retina is still coding the features present in the sparser sequences but is not responding to the additional features of the more dense sequences. We can relate these results to the hypothesis of optimal coding of natural scenes~\citep{Geisler:2008gu}, in which the visual system has a limited capacity to transmit information, it has adapted through evolution to a small subset of all the possible images and optimally code them, discarding the irrelevant information. Thus, the efficiency of the retina to code the stimuli reaches its peak at a sparseness level that would be closer to what the system has evolved to code, and when presented with stimuli containing more signals, it discards the additional information. Much research has been performed to investigate the relationship between natural images and optimal coding, although the focus has been mainly on the spatiotemporal correlations~\citep{Pitkow:2012dh,Rikhye:2015bg}. We have observed that when keeping the spatiotemporal components constant, the modulation of the sparseness of the stimuli has an effect on the retinal response, and more importantly, it allows us to see the level of sparseness beyond which the efficiency does not increase.

It is important to note that these results are the outcome of an interdisciplinary convergence between image processing (to characterize sparseness in natural images), mathematical modeling (for the synthesis of textures) and neurophysiology (for the recordings and their analysis). In particular, in this original framework, neural recordings are not analyzed \emph{post-hoc}, but are instead tuned by the design of parameterized stimuli. In particular, these stimuli are defined from the analysis of natural images. One limit of the neurophysiological study is that we limited ourselves to a simplistic class of textons (Mexican-hat shaped profiles\if\Supp1 , compare to Supplementary Figure~\ref{fig:EUVIP_loggabor}\fi), both for the analysis and synthesis but that ganglion cells in the retina are known to be selective to a wider class of stimulations. However, we believe that this class of stimuli is general enough to characterize a wide range of different cell types. Indeed, by looking at local combinations of events (such as doublets), one could characterize different sub-types of ganglion cells, both static, oriented or even moving, such as to characterize for instance directionally selective ganglion cells. In particular, by manipulating the statistics in the event's matrix $E$, one could target more specifically each of these sub-types. Moreover, while it  was found that such textures would be useful in characterizing the response of retinal neurons, this should prove also very useful to characterize the response of V1 neurons. In particular, it would be essential to tune the relation of ``neighboring edges'' as characterized by how close they are in visual space space but also in orientation and scale, as measured by the so-called ``association field''~\citep{PerrinetBednar15}.

\if\Supp1
\newpage
%⎯⎯⎯⎯⎯⎯⎯⎯⎯⎯⎯⎯⎯⎯⎯⎯⎯⎯⎯⎯⎯⎯⎯⎯⎯⎯⎯⎯⎯⎯⎯⎯⎯⎯⎯⎯⎯⎯⎯⎯⎯⎯⎯⎯⎯⎯⎯⎯⎯⎯⎯⎯⎯⎯⎯⎯⎯⎯⎯⎯⎯⎯⎯⎯⎯⎯⎯⎯⎯⎯⎯⎯⎯⎯⎯⎯⎯⎯⎯⎯⎯⎯
\section{Supplementary material: Sparse coding algorithm}
%⎯⎯⎯⎯⎯⎯⎯⎯⎯⎯⎯⎯⎯⎯⎯⎯⎯⎯⎯⎯⎯⎯⎯⎯⎯⎯⎯⎯⎯⎯⎯⎯⎯⎯⎯⎯⎯⎯⎯⎯⎯⎯⎯⎯⎯⎯⎯⎯⎯⎯⎯⎯⎯⎯⎯⎯⎯⎯⎯⎯⎯⎯⎯⎯⎯⎯⎯⎯⎯⎯⎯⎯⎯⎯⎯⎯⎯⎯⎯⎯⎯⎯
\label{sec:sparse} % \if\Supp1 ; see SI Section~\ref{sec:sparse} for a full description of the algorithm

In general, a greedy approach is applied when finding the best combination is
difficult to solve globally, but can be solved progressively,
one element at a time.
Applied to our problem, the greedy approach corresponds to first choosing
the single filter $\Phi_i$ that best fits the image along with a suitable coefficient $a_i$,
such that the single source $a_i\Phi_i$ is a good match to the image.
Examining every filter $\Phi_j$, we find the filter $\Phi_i$
with the maximal correlation coefficient, where:
\begin{equation}
i = \mbox{argmax}_j \left( \left\langle \frac{\mathbf{I}}{\| \mathbf{I} \|} , \frac{
\Phi_j}{\| \Phi_j\|} \right\rangle \right),
\label{eq:coco}
\end{equation}
$\langle \cdot,\cdot \rangle$ represents the inner product, and $\| \cdot \|$
represents the $\ell_2$ (Euclidean) norm. Since filters at a given scale and orientation
are generated by a translation, this operation can be efficiently computed using a convolution,
but we keep this notation for its generality.
The associated coefficient is the scalar projection:
\begin{equation}
a_{i} = \left\langle \mathbf{I} , \frac{ \Phi_{i}}{\| \Phi_{i}\|^2} \right\rangle
\label{eq:proj}
\end{equation}
Second, knowing this choice, the image can be
decomposed as
\begin{equation}
\mathbf{I} = a_{i} \Phi_{i} + \bf{R}
\label{eq:mp0} \end{equation}
where $\bf{R}$ is the residual image.
We then repeat this 2-step process on the residual (that is, with $\mathbf{I} \leftarrow \bf{R}$)
until some stopping criterion is met.
Note also that the norm of the filters has no influence in this algorithm
on the choice function or on the reconstruction error.
For simplicity and without loss of generality,
we will thereafter set the norm of the filters to $1$: $\forall j, \| \Phi_j \| =1$.
Globally, this procedure gives us a sequential algorithm for reconstructing the signal
using the list of sources (filters with coefficients), which greedily optimizes the $\ell_0$ pseudo-norm
(i.e., achieves a relatively sparse representation given the stopping criterion).
The procedure is known as the Matching Pursuit (MP) algorithm~\citep{Mallat93},
which has been shown to generate good approximations for natural images~\citep{Perrinet10shl}.

For this work we made two minor improvements to this method:
First, we took advantage of the response of the filters as complex numbers.
As stated above, the modulus gives a response independent of the phase of the filter,
and this value was used to estimate the best match of the residual image
with the possible dictionary of filters (Matching step).
Then, the phase was extracted as the argument of the corresponding coefficient
and used to feed back onto the image in the Pursuit step.
This modification allows for a phase-independent detection of edges,
and therefore for a richer set of configurations,
while preserving the precision of the representation.

%------------------------------%
%: \if\Supp1 See Supplementary Figure~\ref{fig:EUVIP_loggabor} \fi
\begin{figure}%[ht!]%[p!]
\centering{\includegraphics[width=\linewidth]{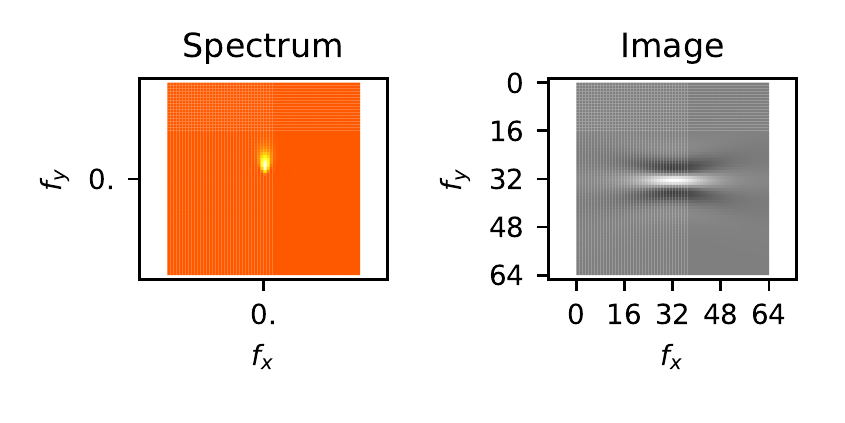}}
\caption{
{\bf Filters used in this paper.} To mimic the average profile of receptive fields in the primary visual cortex, we use localized, orientation-selective filters. These are characterized by a cone-shaped spectrum (Left) and provide with the log-Gabor wavelets (Right)
\label{fig:EUVIP_loggabor}}%
\end{figure}%
%------------------------------%

Second, we used a ``smooth'' Pursuit step.
In the original form of the Matching Pursuit algorithm,
the projection of the Matching coefficient is fully removed from the image,
which allows for the optimal decrease of the energy of the residual
and allows for the quickest convergence of the algorithm
with respect to the $\ell_0$ pseudo-norm
(i.e., it rapidly achieves a sparse reconstruction with low error).
However, this efficiency comes at a cost,
because the algorithm may result in non-optimal representations
due to choosing edges sequentially and not globally.
This is often a problem when edges are aligned (e.g. on a smooth contour),
as the different parts will be removed independently, potentially leading
to a residual with gaps in the line.
Our goal here is not to get the fastest decrease of energy,
but rather to provide a good representation of edges along contours.
We therefore used a more conservative approach,
removing only a fraction (denoted by $\alpha$)
of the energy at each pursuit step (for MP, $\alpha=1$).
We found that $\alpha=0.5$ was a good compromise between rapidity and smoothness.
One consequence of using $\alpha<1$ is that, when removing energy along contours,
edges can overlap; even so, the correlation is invariably reduced.
Higher and smaller values of $\alpha$ were also tested,
and gave classification results similar to those presented here.

In summary, the whole learning algorithm is given by the following nested loops
in pseudo-code:
\begin{enumerate}
\item draw a signal $\mathbf{I}$ from the database; its energy is $E = \| \mathbf{I} \|^2$,
\item initialize sparse vector $\mathbf{s}$ to zero and linear coefficients $\forall j, {a}_j=<\mathbf{I}, \Phi_j >$,
\item while the residual energy $E = \| \mathbf{I} \|^2$ is above a given threshold do:
\begin{enumerate}
\item select the best match: $i = \mbox{ArgMax}_{j} | {a}_j |$, where $| \cdot |$ denotes the modulus,
\item increment the sparse coefficient: $s_{i} = s_{i} + \alpha \cdot {a}_{i}$,
\item update residual image: $ \mathbf{I} \leftarrow \mathbf{I} - \alpha \cdot a_{i} \cdot \Phi_{i} $,
\item update residual coefficients: $\forall j, {a}_j \leftarrow {a}_j - \alpha \cdot a_{i} <\Phi_{i} , \Phi_j > $,
\end{enumerate}
\item the final non-zero values of the sparse representation vector
$\mathbf{s}$, give the list of edges representing the
image as the list of couples $(i, s_{i})$, where $i$ represents an edge occurrence
as represented by its position, orientation and scale.
\end{enumerate}
%Note that the correlations $<\Phi_{i^\ast} , \Phi_i >$ are known \emph{a
						%priori} laurent: TODO say we have lots of controls:
						%more N, more theta, smooth vs hard

%------------------------------%
%: \if\Supp1 See Supplementary Figure~\ref{fig:EUVIP_lena_movie} \fi
\begin{figure}%[ht!]%[p!]
\centering{\includegraphics[width=\linewidth]{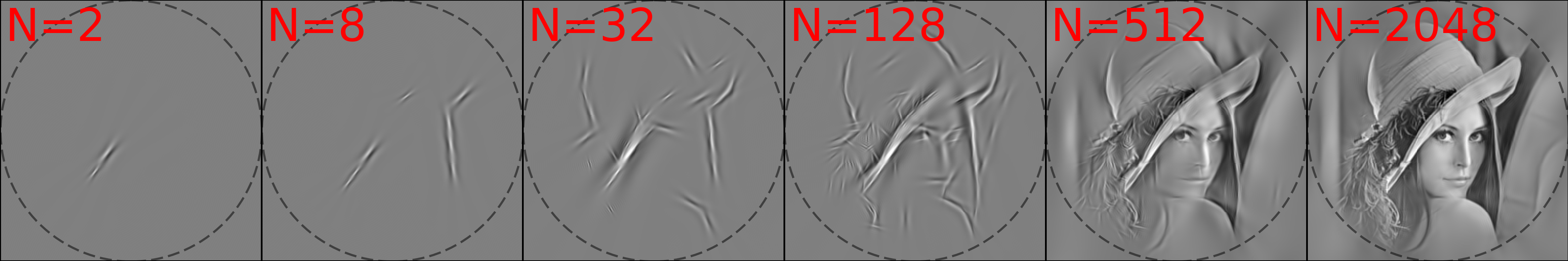}}
\caption{
{\bf Progressive reconstruction as a function of sparseness.}
\label{fig:EUVIP_lena_movie}}%
\end{figure}%

\fi

\bibliographystyle{plainnat}
\bibliography{Perrinet16EUVIP}
\end{document}